\documentclass[aip,graphicx]{revtex4-1}

\usepackage{graphicx}
\usepackage{dcolumn}
\usepackage{bm}
\usepackage{subfigure}
\draft 

\begin{document}


\title{Sr and Mn co-doped LaCuSO: a wide band gap oxide diluted magnetic semiconductor with $T_{C}$ around 200 K} 



\author{Xiaojun Yang}
\affiliation{Department of Physics and State Key Lab of Silicon Materials, Zhejiang University, Hangzhou 310027, China}
\author{Yuke Li}
\affiliation{Department of Physics and State Key Lab of Silicon Materials, Zhejiang University, Hangzhou 310027, China}
\affiliation{Department of Physics, Hangzhou Normal University, Hangzhou 310036, China}
\author{Chenyi Shen}
\author{Bingqi Si}
\author{Yunlei Sun}
\author{Qian Tao}
\author{Guanghan Cao}
\author{Zhuan Xu\footnote[1]{Electronic address: zhuan@zju.edu.cn}}
\author{Fuchun Zhang}
\affiliation{Department of Physics and State Key Lab of Silicon Materials, Zhejiang University, Hangzhou 310027, China}


\date{\today}

\begin{abstract}
Here we report the synthesis of a bulk oxide diluted magnetic
semiconductor (DMS) system
La$_{1-x}$Sr$_{x}$Cu$_{0.925}$Mn$_{0.075}$SO (\emph{x} = 0, 0.025,
0.05, 0.075 and 0.1). As a wide band gap p-type oxide
semiconductor, LaCuSO satisfies all the conditions forecasted
theoretically to be a room temperature DMS. The Curie temperature
($T_{C}$) is around 200 K as \emph{x} $\geq$ 0.05, which is among
the highest $T_{C}$ record of known bulk DMS materials up to now.
The system provides a rare example of oxide DMS system with p-type
conduction, which is important for formation of high temperature
spintronic devices.
\end{abstract}


\maketitle 

The discovery of ferromagnetism in Mn-doped GaAs has drawn
researchers to the field of Diluted Magnetic Semiconductor (DMS),
which makes it possible to examine tremendous phenomena such as
quantum Hall effects, semiconductor lasers and single-electron
charging, and could bring about numerous applications in sensors
and memories as well as computing with electron spins.\cite{1ohno
Science,2I. Zutic RMP,3T. Dietl Science,4T. Dietl Nat. Mat.,5 H.
Ohno APL} However, the most applications are restricted by the low
ferromagnetic transition temperature of III-V semiconductor based
DMS.\cite{6} Followed by the observation of ferromagnetism in
dilutely cobalt doped TiO$_{2}$,\cite{7} much attention was paid
to oxide DMS systems, and room-temperature ferromagnetism was
realized in oxide DMS systems, \cite{8,9,10,11} though the field
has not yet reached the same level of maturity and clarity about
the attendant phenomena as the III-V semiconductor based DMS.
Owing to a limited number of p-type oxide systems are available
and the difficulties encountered in converting natural n-type
oxides such as ZnO to p-type conduction by doping, there have been
only a limited number of studies on p-type oxide DMS systems.
\cite{10,11} More work on p-type materials should be interesting,
and p-type wide band gap oxide DMS systems are important for
formation of spintronics devices.

Due to the limitation of the chemical solubility of Mn, most of
the DMS systems are chemically metastable, and available only as
thin films. Accordingly, the quality of the thin film material
depends sensitively on the preparation methods and heat
treatments.\cite{12}  Only the bulk crystalline specimens make it
possible to perform  muon spin relaxation ($\mu$SR), nuclear
magnetic resonance (NMR) and even neutron scattering. Therefore
the bulk DMS materials are highly required to obtain more reliable
results. \cite{13} Recently, a Mn doped pnictide Li(Zn,Mn)As was
reported to be a bulk DMS material with a Curie temperature
($T_{C}$) of about 50 K.\cite{15} Layer-structured pnicitides
thus become a promising avenue for exploring bulk DMS materials.

LaCuSO was initially prepared through oxidation of LaCuS$_{2}$,
and its crystal structure was determined as the tetragonal
ZrCuSiAs-type (also called 1111 type), \cite{16,17,18} which is
identical to that of LaFeAsO, a typical parent compound of the
1111 type iron-based pnictide superconductor.\cite{19,20,21}
LaCuSO is a wide band gap (3.1eV) p-type conductive oxysulfides
semiconductor,\cite{18} and it can become metallic with Sr
doping.\cite{23} Apparently, as a wide band gap p-type oxide
semiconductor, LaCuSO is an attractive candidate for p-type oxide
DMS. There are a few theoretical studies on Mn doped
LaCuSO,\cite{theory_Mn_LaCuSO} but only the samples with low Mn
doping level were experimentally synthesized, and no
ferromagnetism was found,\cite{(Ca_Mn)LaCuSO} which should be due
to the low Mn concentration and low charge carrier density.
Moreover, LaCuSO is important to transparent p-n junctions and
optoelectronic devices.\cite{27,LaCuSO_film}

Here we report the synthesis of a bulk p-type oxide DMS in the Sr
and Mn co-doped La$_{1-x}$Sr$_{x}$Cu$_{0.925}$Mn$_{0.075}$SO
(\emph{x} = 0, 0.025, 0.05, 0.075 and 0.1) system.
The La$_{1-x}$Sr$_{x}$Cu$_{0.925}$Mn$_{0.075}$SO
(\emph{x} = 0, 0.025, 0.05, 0.075 and 0.1) system exhibits ferromagnetic order with
\emph{T}$_{C}$ around 200 K, which is among
the highest $T_{C}$ record of known bulk DMS materials up to now.\cite{7,8,9,10,11,BaZn2As2,GaAs_high Tc,6_173K}
By contrast, due to the low
solubility limit of Mn ($<$0.5 mol \%), Mn doped LaCuSeO does not
exhibit intrinsic ferromagnetism \cite{13}. As a wide band gap
p-type conductive oxysulphide semiconductor, LaCuSO satisfies all
the conditions forecasted by Dietl \emph{et al}. to be a room
temperature DMS. \cite{3T. Dietl Science,13} We expect that the
solubility of Sr and Mn can be raised to a higher level with
epitaxy and/or nanotechnology, and a room temperature p-type oxide
DMS may be realized in thin films and/or nanostructured materials.
Our discovery makes the formation of high temperature spintronics
p-n junction devices become possible.

The polycrystalline samples with nominal formula
La$_{1-x}$Sr$_{x}$Cu$_{0.925}$Mn$_{0.075}$SO (\emph{x} = 0, 0.025,
0.05, 0.075 and 0.1) were synthesized by solid state reaction
method.\cite{Solid state} All the starting material, La ingot, the
powders of La$_{2}$O$_{3}$, Cu, Mn, S and SrS are of high purity
(¡Ý99.9\%). First, the La ingot was grounded to powder by hand.
The powders of these materials were weighted according to the
stoichiometric ratio, and then thoroughly mixed in an agate
mortar. The mixtures were pressed into pellets under a typical
pressure of 2000 kg cm$^{-2}$. All these processes were operated
in a glove box filled with high-purity argon. The pellets were put
into crucibles and sealed in evacuated quartz tubes, then heated
up with a ramping rate of 0.5 K/min to 1223 K and kept at that
temperature for 2000 min, and finally furnace cooled to room
temperature.

Powder x-ray diffraction (XRD) was performed at room temperature using a PANalytical x-ray diffractometer (Model EMPYREAN) with a monochromatic CuK$_{\alpha1}$ radiation. The XRD diffractometer system was calibrated using standard Si powders. The detailed structural parameters were obtained by Rietveld refinement using step-scan (4 s/step) data. The dc magnetization measurement was carried out on a Quantum Design magnetic property measurement system (MPMS-5) employing both zero-field-cooling (ZFC) and field-cooling (FC) protocols. For the measurement of isothermal magnetization, the magnetic field was applied in a field sweep mode. The thermopower was measured by a steady-state technique.

\begin{figure}
\includegraphics[width=8cm]{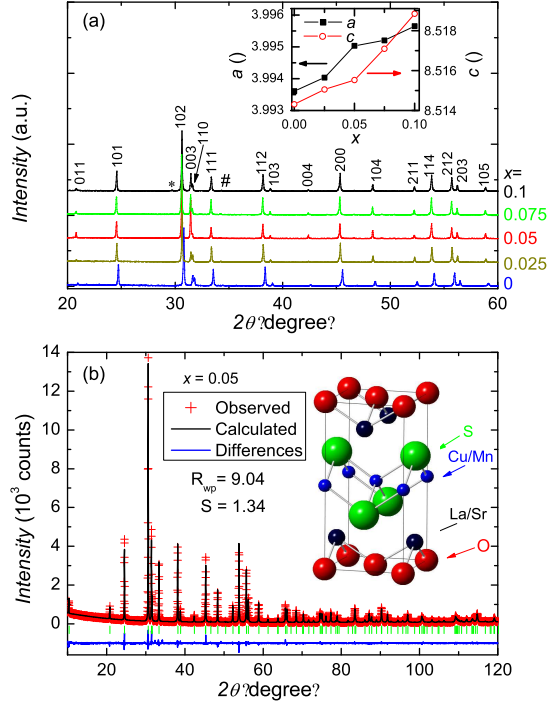}
\caption{\label{Fig. 1} (a), (color line) Room-temperature XRD
patterns of La$_{1-x}$Sr$_{x}$Cu$_{0.925}$Mn$_{0.075}$SO (\emph{x}
= 0, 0.025, 0.05, 0.075 and 0.1) compounds. $\ast$, $\dag$ and
$\sharp$ denote the minor peaks due to impurity phases SrS,
(La,Sr)MnO$_{3}$ and MnS (Alabandite), respectively. The impurity
phase of (La,Sr)MnO$_{3}$ can only be detected for \emph{x} = 0.1
sample (indicated by $\dag$). Inset of (a) exhibits the variation
of lattice constants a (squares) and c (circles) with Sr doping
content ($x$). (b), Rietveld refinement of the powder X-ray
diffraction for the \emph{x} = 0.05 sample. Inset of (b) shows the
crystal structure of (La,Sr)(Cu,Mn)SO belonging to the tetragonal
ZrCuSiAs-type structure.}
\end{figure}

Figure 1(a) displays the X-ray diffraction patterns of La$_{1-x}$Sr$_{x}$Cu$_{0.925}$Mn$_{0.075}$SO for \emph{x} = 0, 0.025, 0.05, 0.075 and 0.1, respectively, and Fig. 1(b) shows a representative Rietveld refinement of the sample with \emph{x} = 0.05.  All the peaks are assigned the same as those for the LaCuSO phase and can be well indexed based on the P4/nmm (No. 129) space group with tetragonal ZrCuSiAs-type structure, except a few minor peaks assigned as impurity phases, which indicates that the samples are mostly composed of single phases (more than 99\%). For \emph{x} = 0.1, a trace of (La,Sr)MnO$_{3}$ can be detected (indicated by $\dag$ in Fig.1(a)). More information on the solubility of Mn in LaCuSO can be found in the supplementary material.\cite{SM} The lattice parameters of the samples were obtained by Rietveld refinement, as shown in the inset of Fig. 1(a), both the $c$-axis and $a$-axis expand slightly with increasing Sr content (\emph{x}). This should be due to the slightly larger ionic radius of Sr$^{2+}$ compared with La$^{3+}$. The Rietveld refinement of \emph{x} = 0.05 sample based on the ZrCuSiAs-type structure shows that the calculated profile well matches the experimental data. The weighted reliable factor $R_{wp}$ and the goodness of fit $S$ are 9.04\% and 1.34, respectively, indicating a good refinement. The inset of Fig. 1(b) exhibits the crystal structure of LaCuSO which belongs to tetragonal ZrCuSiAs-type structure.

\begin{figure}
\includegraphics[width=8cm]{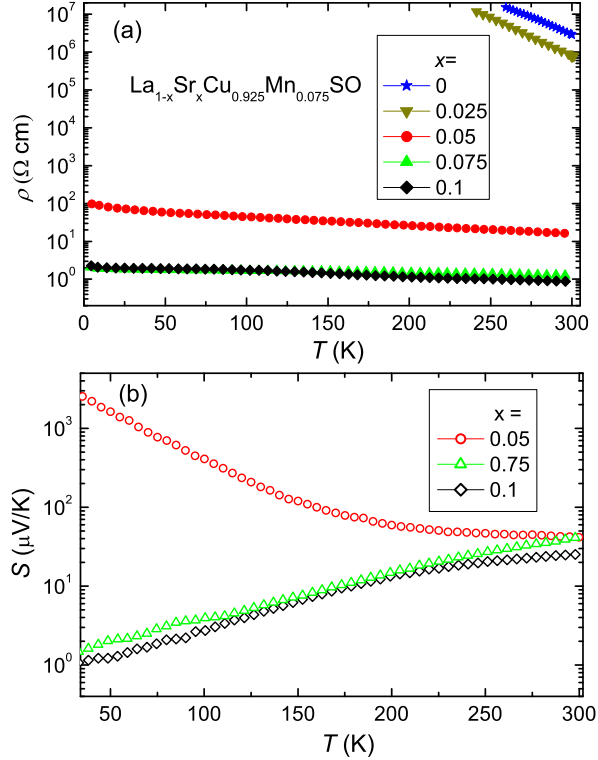}
\caption{\label{Fig. 2} (a), Temperature dependence of electrical resistivity of La$_{1-x}$Sr$_{x}$Cu$_{0.925}$Mn$_{0.075}$SO (\emph{x} = 0, 0.025, 0.05, 0.075 and 0.1) specimens. (b) Temperature dependence of thermopower for the \emph{x} = 0.05, 0.075 and 0.1 samples.}
\end{figure}

The temperature dependence of electrical resistivity of
La$_{1-x}$Sr$_{x}$Cu$_{0.925}$Mn$_{0.075}$SO (\emph{x} = 0, 0.025,
0.05, 0.075 and 0.1) specimens is shown in Fig. 2(a). We used
bar-shaped samples of typical sizes 2 $\times$ 1 $\times$ 0.5 mm
for transport property measurements. At least two sets of samples
were checked for each doping level, and the results are consistent
with each other within an error of 10\%. The \emph{x} = 0 sample
is highly insulating, whose electrical resistivity ($\rho$) is as
large as 2.8 M$\Omega$ cm in the room temperature, and it
increases drastically with decreasing temperature (it increases
beyond our measurement limitation below 260 K). Other transport
properties such as Hall and Seebeck coefficients are also
difficult to measure due to the extremely low carrier density.
With Sr doping, hole-type charge carrier can be introduced and
resistivity decreases rapidly. Electrical resistivity of all the
\emph{x} = 0.05, 0.075 and 0.1 samples exhibits typical
semiconductor behavior as well, i.e., it increases rapidly with
decreasing temperature. As it will be shown below, the samples
with a higher Sr doping level exhibit ferromagnetism with $T_{C}$
as high as 199 K. Namely, the \emph{x} = 0.05, 0.075 and 0.1
samples are p-type oxide DMS. Fig. 2(b) displays the thermopower
of $x$ = 0.05, 0.075 and 0.1 samples. The positive thermopower
indicates that p-type electrical conduction is dominant in these
samples. The thermopower of \emph{x} = 0.05 samples increases with
decreasing temperature, consistent with the typical behavior of a
semiconductor. With more Sr doping, the thermopower decrease
drastically. The thermopower of $x$ = 0.075 and 0.1 samples
decreases with decreasing temperature, which deviates from the
typical semiconductor behavior.

\begin{figure}
\includegraphics[width=8cm]{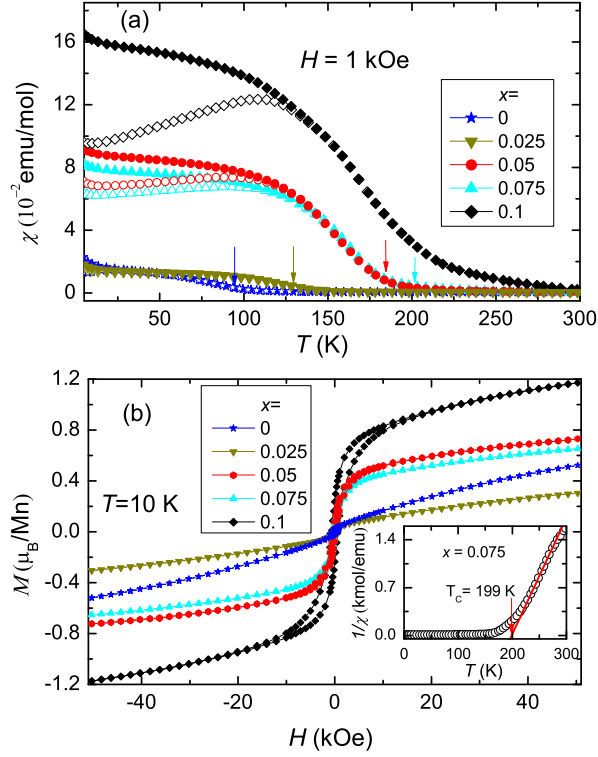}
\caption{\label{Fig. 3}(a),  Temperature dependence of dc magnetic
susceptibility measured under $\emph{H}$ = 1 kOe for the
La$_{1-x}$Sr$_{x}$Cu$_{0.925}$Mn$_{0.075}$SO (\emph{x} = 0, 0.025,
0.05, 0.075 and 0.1) specimens, with solid symbols standing for
field-cooling (FC) and open ones for zero-field-cooling(ZFC). (b),
Field dependence of magnetization measured at 10 K for
La$_{1-x}$Sr$_{x}$Cu$_{0.925}$Mn$_{0.075}$SO (\emph{x} = 0, 0.025,
0.05, 0.075 and 0.1) samples. Inset: Plot of 1/$\chi$ vs. $T$ and
the determination of the Curie temperature $T_C$ is shown.}
\end{figure}

Figure 3 (a) shows the temperature dependence of dc magnetic
susceptibility in zero-field-cooling (ZFC) and field-cooling (FC)
procedures under $H$ = 1 kOe for
La$_{1-x}$Sr$_{x}$Cu$_{0.925}$Mn$_{0.075}$SO (\emph{x} = 0, 0.025,
0.05, 0.075 and 0.1) samples, with solid symbols standing for FC
and open ones for ZFC. Apparently the samples become
ferromagnetically ordered. Denoted by arrows, the Curie
temperatures  ($T_{C}$), which are defined as the x-axis intercept
of the linear fitting curves of the temperature dependence of
inverse magnetic susceptibility 1/$\chi$ (see the inset of Fig.
3(b)), are 92, 130, 180 and 199 K for \emph{x} = 0, 0.025, 0.05
and 0.075 samples, respectively.  When the Sr doping level
(\emph{x}) reaches 0.1, a long tail above 200 K can be observed,
which may result from a tiny impurity phase (La,Sr)MnO$_{3}$. The
Rietveld analysis shows that the impurity phase (La,Sr)MnO$_{3}$
is less than 0.3\%. (La,Sr)MnO$_{3}$ is a ferromagnetic compound
with $T_{C}$ of about 300 K. \cite{28} This impurity phase cannot
be detected in the XRD patterns of the
La$_{1-x}$Sr$_{x}$Cu$_{0.925}$Mn$_{0.075}$SO  samples with low Sr
content (\emph{x} $<$ 0.1). But it can be clearly found in the
samples with either high Sr content or high Mn doping level (such
as 0.1) specimens. $T_{C}$ of the \emph{x} = 0.1 sample appears to
be over 200 K. As shown in the inset of Fig. 3(b), a small
deviation of linearity in 1/$\chi$ above 280 K is probably due to
the presence of trace amount of ferromagnetic (La,Sr)MnO$_{3}$
impurity. The $T_{C}$ increases with the Sr doping level
(\emph{x}),which should be proportional to the carrier density,
indicating that the induction of hole-type charge carriers by Sr
doping is crucial to the enhancement of FM order and that is
consistent with the carrier-induced origin of the ferromagnetism.
The magnetic hysteresis loop $M(H)$ curves of
La$_{1-x}$Sr$_{x}$Cu$_{0.925}$Mn$_{0.075}$SO ($\emph{x}$ = 0,
0.025, 0.05, 0.075 and 0.1) specimens at $\emph{T}$ = 10 K are
shown in Fig. 3(b). Although the ferromagnetic hysteresis is
small, the $M(H)$ curves exhibit typical FM behavior. The
saturating magnetic moments reach 0.52, 0.30, 0.73, 0.65 and 1.17
$\mu$$_{B}$ per Mn atom at $\emph{H}$ = 50 kOe, for \emph{x} = 0,
0.025, 0.05, 0.075 and 0.1 samples, respectively, which are
comparable with those in (Ga, Mn)As\cite{1ohno Science} and Li(Mn,
Zn)As\cite{15}. The observed magnetic moment strongly supports the
bulk ferromagnetism and rules out the possibility that the
magnetism is due to the impurity phase of (La, Sr)MnO$_{3}$, whose
mass fraction is as small as 0.3\%(obtained by Rietveld
refinement) for $x$ = 0.1 and even undetectable for $x$ $<$ 0.1.
For the sample with $x$ = 0.1, the saturating magnetization is
0.13 $\mu_{B}$/Mn at $T$ = 200 K(not shown here) after subtracting
the small $T$-linear contribution. The obtained saturating
magnetic moments consistent with the fact that the mass fraction
of (La, Sr)MnO$_{3}$ is as small as 0.3\%, assuming the ordered
moment in (La, Sr)MnO$_{3}$ is 3 $\sim$ 4 $\mu_{B}$ per Mn
atom.\cite{28}

A small $H$-linear component can be found in the \emph{M(H)}
curves which may result from the remaining paramagnetic spins
and/or field-induced polarization. Moreover, a divarication
between the ZFC and FC curves is observed. Usually such a
divarication can result from the magnetic hysteresis of FM ordered
state. However, a magnetic glass state could have a similar
divarication between the ZFC and FC curves.\cite{29,30} The
magnetic behavior demonstrated here should not result from the
canonical spin glass (SG) systems since the samples still carry
large magnetic moment down to 5 K.

In summary, a wide band gap p-type bulk oxide DMS system
La$_{1-x}$Sr$_{x}$Cu$_{0.925}$Mn$_{0.075}$SO has been synthesized.
The bulk specimens with suitable Sr and Mn doping (i.e., \emph{x}
= 0.05, 0.075) exhibit ferromagnetic order with $T_{C}$ around 200
K. The induction of hole-type charge carriers by Sr doping is
crucial to the enhancement of FM order. LaCuSO satisfies all the
conditions forecasted by Dietl \emph{et al}.\cite{3T. Dietl
Science} for realizing room temperature DMS. It is expected that
the solubility of Sr and Mn can be raised to a higher level with
epitaxy and/or nanotechnology, and a room temperature DMS may be
realized in thin films and/or nanomaterials. Our discovery makes
the formation of high temperature spintronics p-n junction devices
possible.

 We acknowledge the support from the National Basic Research Program of China (Grant Nos. 2011CBA00103 and 2012CB821404), NSF of China (Contract Nos. 11174247 and 11190023), and the Fundamental Research Funds for the Central Universities of China.


%
%

%




\begin{thebibliography}{00}

\bibitem{1ohno Science}  H. Ohno, Science \textbf{281}, 951 (1998).
\bibitem{2I. Zutic RMP}   I. \v{Z}uti\'{c}, J. Fabian, and S. Das Sarma, Rev. Mod. Phys. \textbf{76}, 323 (2004).
\bibitem{3T. Dietl Science}   T. Dietl, H. Ohno, F. Matsukura, J. Cibert, and D. Ferrand, Science  \textbf{287}, 1019 (2000).
\bibitem{4T. Dietl Nat. Mat.}  T. Dietl, Nature Mater.  \textbf{9}, 965 (2010).
\bibitem{5 H. Ohno APL}   H. Ohno, A. Shen, F. Matsukura, A. Oiwa, A. Endo, S. Katsumoto, and Y. Iye, Appl. Phys. Lett. \textbf{69}, 363 (1996).
\bibitem{6}   D. Chiba, K. Takamura, F. Matsukura, and H. Ohno, Appl. Phys. Lett.  \textbf{82}, 3020 (2003).
\bibitem{7}   Y. Matsumoto, M. Murakami, T. Shono, T. Hasegawa, T. Fukumura, M. Kawasaki, P. Ahmet, T. Chikyow, S. Koshihara, and H. Koinuma, Science \textbf{291}, 854 (2001).
\bibitem{8} S. B. Ogale, Adv. Mater.  \textbf{22}, 3125 (2010).
\bibitem{9}  K. A. Griffin, A. B. Pakhomov, C. M. Wang, S. M. Heald, and K. M. Krishnan, Phys. Rev. Lett. \textbf{ 94}, 157204 (2005).
\bibitem{10}  K. R. Kittilstved, N. S. Norberg, and D. R. Gamelin, Phys. Rev. Lett.  \textbf{94}, 147209 (2005).
\bibitem{11}  J. M. D. Coey, M. Venkatesan, and C. B. Fitzgerald, Nature Mater.  \textbf{4}, 173 (2005).
\bibitem{12}  S. J. Potashnik, K. C. Ku, S. H. Chun , J. J. Berry, N. Samarth, and P. Schiffer, Appl. Phys. Lett.  \textbf{79}, 1495 (2001).
\bibitem{13}  H. Yanagi, S. Ohno, T. Kamiya, H. Hiramatsu, M. Hirano, and H. Hosono,  J. Appl. Phys.  \textbf{100}, 033717 (2006).
\bibitem{15}   Z. Deng, C. Q. Jin, Q. Q. Liu, X. C. Wang, J. L. Zhu, S. M. Feng, L. C. Chen, R. C. Yu, C. Arguello, T. Goko, F. Ning, J. Zhang, Y. Wang, A. A. Aczel, T. Munsie, T. J. Williams, G. M. Luke, T. Kakeshita, S. Uchida, W. Higemoto, T. U. Ito, Bo Gu, S. Maekawa, G. D. Morris, and Y. J. Uemura, Nat. Commun.  \textbf{2}, 422 (2011).
\bibitem{16}  M. Palazzi, C. R. Acad. Sci., Paris, S\'{e}r. I \textbf{292}, 789 (1981).
\bibitem{17}  K. Ueda, S. Inoue, S. Hirose, H. Kawazoe, and H. Hosono, Appl. Phys. Lett.  \textbf{77}, 2701 (2000).
\bibitem{18}  S. Inoue, K. Ueda, H. Hosono, and N. Hamada, Phys. Rev. B  \textbf{64}, 245211 (2001).
\bibitem{19}  Y. Kamihara, T. Watanabe, M. Hirano, and H. Hosono, J. Am. Chem. Soc.  \textbf{130}, 3296 (2008).
\bibitem{20}  X. H. Chen, T. Wu, G. Wu, R. H. Liu, H. Chen, and D. F. Fang, Nature  \textbf{453}, 761 (2008).
\bibitem{21}  C. Wang, L. J. Li, S. Chi, Z. W. Zhu, Z. Ren, Y. K. Li, Y. T. Wang, X. Lin, Y. K. Luo, S. Jiang, X. F. Xu, G. H. Cao, and Z. A. Xu, Europhys. Lett.  \textbf{83}, 67006 (2008).
\bibitem{23}  Y. Takano, K. Yahagi, and K. Sekizawa, Physica B  \textbf{206 and 207}, 764 (1995).
\bibitem{theory_Mn_LaCuSO} V. V. Bannikov, and A. L. Ivanovskii, Phys. Solid State \textbf{54}, 1117 (2012).
\bibitem{(Ca_Mn)LaCuSO} K. Takase, O. Shoji, T. Shimizu, Y. Takahashi, Y. Takano, and K. Sekizawa, J. Magn. Magn. Mater. \textbf{272-276}, E1535 (2004).
\bibitem{27}  K. Ueda, S. Inoue, H. Hosono,N. Sarukura, and M. Hirano, Appl. Phys. Lett.  \textbf{78}, 2333  (2001).
\bibitem{LaCuSO_film} H. Hiramatsua, K. Ueda, H. Ohta, M. Orita, M. Hirano, and H. Hosono, Thin Solid Films \textbf{411}, 125 (2002).
\bibitem{BaZn2As2} K. Zhao, Z. Deng, X. C. Wang, W. Han, J. L. Zhu, X. Li, Q. Q. Liu, R. C. Yu, T. Goko, B. Frandsen, L. Liu, F. Ning, Y. J. Uemura, H. Dabkowska, G. M. Luke, H. Luetkens, E. Morenzoni, S. R. Dunsiger, A. Senyshyn, P. Boni, and C. Q. Jin, Nat. Commun. \textbf{4}, 1442 (2013).
\bibitem{GaAs_high Tc} K. Olejn\'{i}k, M. H. S. Owen, V. Nov\'{a}k, J. Ma\v{s}ek, A. C. Irvine, J. Wunderlich, and T. Jungwirth, Phys. Rev. B \textbf{78}, 054403 (2008).
\bibitem{6_173K} T. Jungwirth, K. Y. Wang, J. Ma\v{s}ek, K. W. Edmonds, J. K\"{o}nig, J. Sinova, M. Polini, N. A. Goncharuk, A. H. MacDonald, M. Sawicki, A. W. Rushforth, R. P. Campion, L. X. Zhao, C. T. Foxon, and B. L. Gallagher, Phys. Rev. B \textbf{72}, 165204 (2005).
\bibitem{Solid state} S. Koyano, K. Takase, Y. Kuroiwa, S. Aoyagi, O. Shoji, K. Sato,
Y. Takahashi, Y. Takano, and K. Sekizawa, J. Alloys Compd. \textbf{408}, 95 (2006).
\bibitem{SM} See supplementary material at [URL will be inserted by AIP] for the information on the solubility of Mn in LaCuSO.
\bibitem{28}  A. Urushibara, Y. Moritomo, T. Arima, A. Asamitsu, G. Kido, and Y. Tokura, Phys. Rev. B  \textbf{51}, 14103  (1995).
\bibitem{29}  X. F. Liu, S. Matsuishi, S. Fujitsu, and H. Hosono, Phys. Rev. B  \textbf{84}, 214439 (2011).
\bibitem{30} X. F. Liu, S. Matsuishi, S. Fujitsu, T. Ishigaki, T. Kamiyama, and H. Hosono, J. Am. Chem. Soc.  \textbf{134}, 11687 (2012).


\end{thebibliography}
\end{document}